\documentclass[prc,aps,nofootinbib,showkeys,showpacs,twocolumn]{revtex4-2}   

\usepackage[T1]{fontenc} %
\usepackage{epsfig}  
\usepackage{graphicx}  
\usepackage{amssymb}
\usepackage{amsmath}
\usepackage[english]{babel}
\usepackage{color}  
\definecolor{darkgreen}{rgb}{0.2,0.5, 0.2}

\usepackage{hyperref}  

\usepackage{lipsum}

\usepackage{dsfont}

\begin{document}
 
\title{ Microscopic Study of Spin Transfer  in Near-Barrier Nuclear Reactions }

\author{Guillaume Scamps}
\affiliation{Laboratoire des 2 Infinis -Toulouse (L2IT-IN2P3), Université de Toulouse, CNRS, UPS, F-31062 Toulouse Cedex 9, France}
  \date{\today}

\begin{abstract}

\begin{description}
\item[Background] 
In quasi-fission, it is unclear what the interaction between the relative orbital angular momentum and the spin of the fragments is from a microscopic perspective. In macroscopic approaches, it is expected that the large value of the relative orbital angular momentum, of the order of 100~$\hbar$ is transferred through tangential dissipation to the fragments' intrinsic spin by sliding and rolling friction.
 
\item[Purpose]  

The goal is to investigate the angular momentum transfer from the initial relative orbital angular momentum to the fragments' spinand to clarify how the transferred spin is distributed between the fragments.  The investigation further aims to establish the timescales associated with different transfer mechanisms and to determine the influence of deformation on this process.
 
\item[Method] 

A time-dependent density functional the-
ory(TDDFT) simulation in the time-dependent Hartree-Fock (TDHF)-Skyrme framework is used to describe several reactions at different impact parameters with increasing complexity. A method is proposed to study the evolution of the fragments' total spin as a function of time and angular velocity.  
 
\item[Results]  
The increasingly complex reaction allows for careful analysis of all the mechanisms responsible for the transfer of spin. In particular, it is shown that the transfer of nucleons, and neck formation can significantly affect the transfer of spin through tangential friction.

\item[Conclusions] 
Several mechanisms are in contradiction with previous macroscopic calculations. In particular, the spin of the fragments does not always increase during the collision which prevents it from being used to estimate the collision time of the reaction.

\end{description}

\end{abstract}

\maketitle   

 \vspace{0.5cm}
 
 \section{Introduction}
 
 The discovery that the fission fragments' spins are not correlated~\cite{Wilson:2021}
 generated important renewal of interest for the generation and correlation of fission fragments spin arise in both theory~\cite{Randrup:2021,Vogt:2021,Marevic:2022,Bulgac:2022b,Scamps:2022,Scamps:2023a,Piau2023,Scamps:2023,Scamps2024} and experience \cite{Francheteau2024,Marin2024}.
 However, these studies assume zero or a negligible total spin of the fissioning system. This simplifies the picture and is a valid assumption in a  few cases, in particular, the spontaneous fission of even-even nuclei.
 In the case of a non-zero total spin, for instance, in quasi-fission reactions, the main question is \textit{ how the total angular momentum is shared between the fragments and the relative orbital angular momentum?}. 
 
 In quasi-fission, the transfer mechanism between the relative orbital angular momentum and the spin of the fragments is expected to be the tangential dissipation. The collective motion describing the rotation of the fragments' center of mass is expected to experience friction forces~\cite{Bass1980,Tsang1974,Back1990,JORGENSEN1987,Wash2020,Ayik2020,VOLKOV1978,min1980transfert}, transferring this angular momentum to the internal angular momentum.  Two types of friction force (\emph{sliding} and \emph{rolling}) corresponding to two different mechanisms are expected to be at play leading to two types of equilibrium. One where the whole di-nuclear system rotates rigidly (\emph{sticking condition}) and one where one fragment "roll" over the other canceling the currents at the neck (\emph{rolling condition})~\cite{Bass1980}.
 
 It is also important to understand the tangential dissipation since this mechanism could help distinguish between the quasi-fission and fusion-fission reactions as proposed in Ref. \cite{Vardaci2020}. If the angular momentum dissipation is slow, a larger spin could be found in fragments during fusion-fission than quasi-fission reactions. The time scale associated with the sticking and rolling frictions can be estimated in the same way as in Ref.~\cite{simenel2020timescales}, where it was found a time scale of about 1~zs for the angular momentum dissipation. The goal of the present article is to investigate extensively the mechanism responsible for the transfer of spin between the fragments.

The time-dependent Hartree-Fock (TDHF) approach~\cite{Negele:1982,simenel2012nuclear} naturally incorporates one-body dissipation through the interaction of nucleons with the collective mean-field potential. A reasonable question is whether this treatment is sufficient or if two-body dissipation is necessary for a more realistic description. While Time-Dependent Density Matrix (TDDM) calculations~\cite{gong1990application,wen2018two} include two-body dissipation, they require significantly more computational resources. Another two-body effect neglected in TDHF is superfluidity, which is expected to act as a lubricant in large-amplitude collective motion such as fission~\cite{Scamps:2015,Tanimura:2015,Bulgac:2019c}, where pairing interactions reduce Landau-Zener dissipation~\cite{landau1932theorie,zener1932non}. Nevertheless, TDHF remains a valuable tool for studying dissipation microscopically (see, for instance, Ref.\cite{washiyama2009one,dai2014dissipation}), since pairing is expected to be quickly suppressed in reactions, making one-body dissipation dominant over two-body dissipation. Indeed, TDDM calculations\cite{wen2018two} show that two-body collisions increase the friction coefficient by only 20 $\%$ in fusion reactions.

Using the microscopic TDHF model, calculations are done for the $^{40}$Ca~+~$^{40}$Ca, $^{208}$Pb~+~$^{208}$Pb, $^{40}$Ca~+~$^{208}$Pb and $^{50}$Ca~+~$^{176}$Yb reactions at energies above the Coulomb barrier. They are chosen in this study to introduce progressive complexity with asymmetry between the initial nuclei and shape deformation in the ground state of the initial fragments.

\begin{figure*}[t]
    \centering
    \includegraphics[height=.46\linewidth]{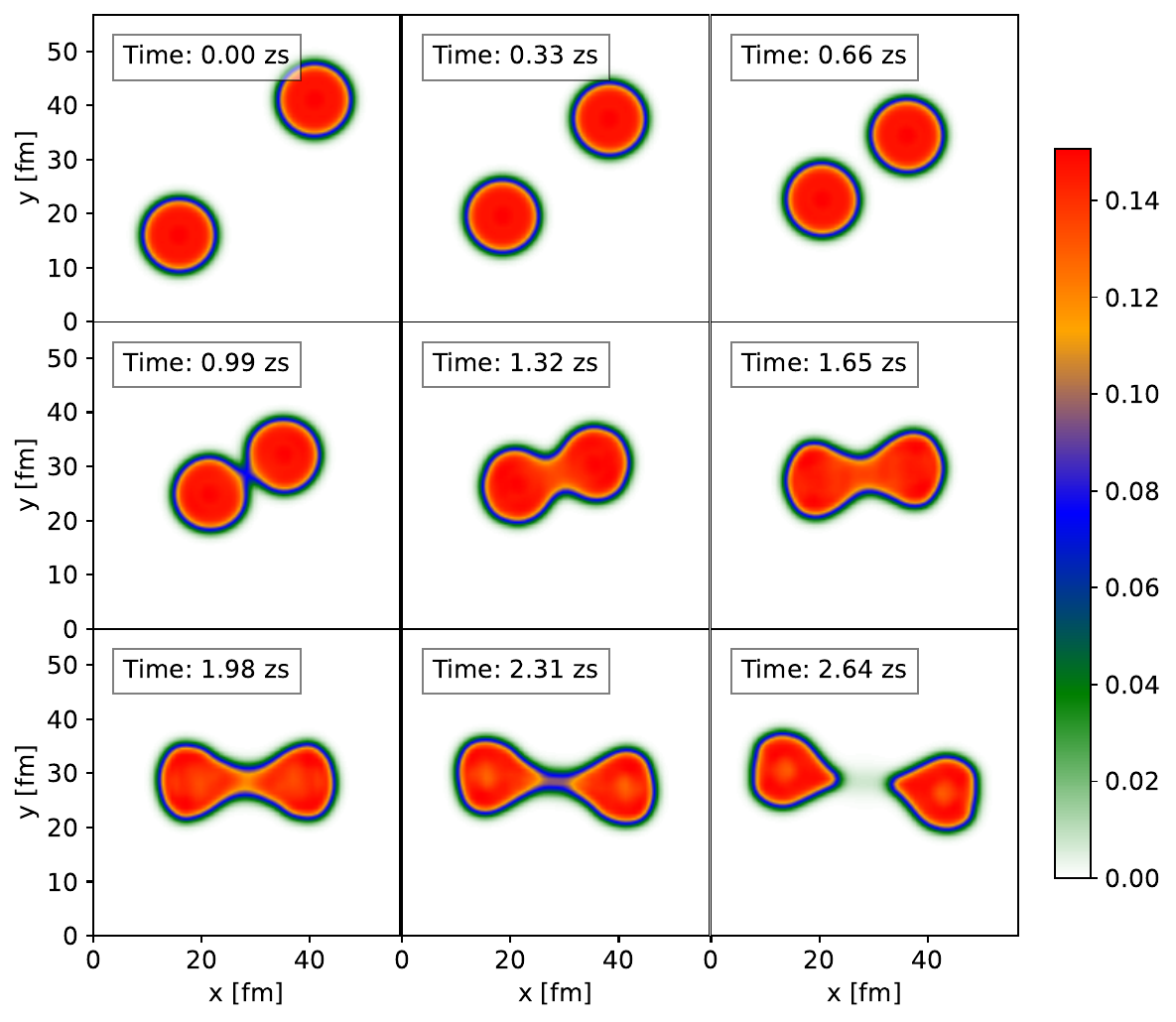}
    \includegraphics[height=.46\linewidth]{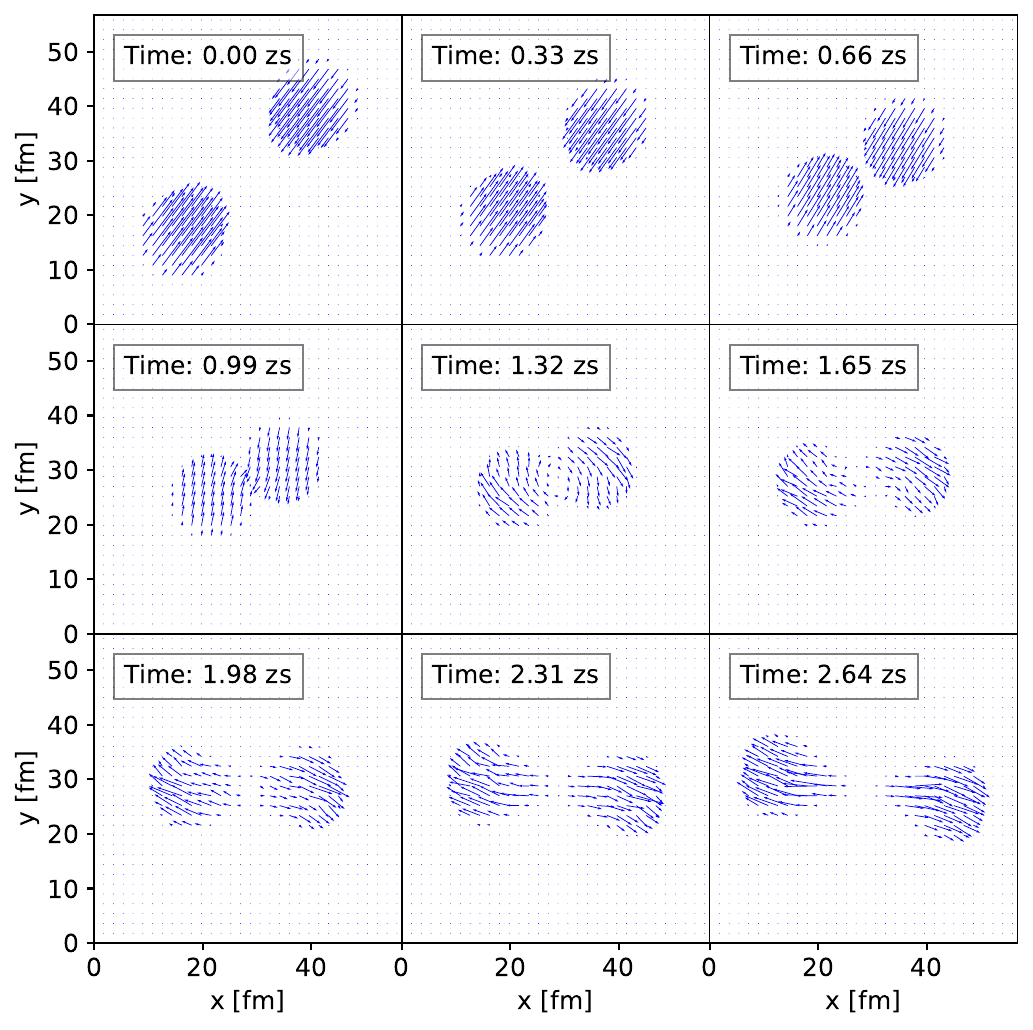}
    \caption{Evolution of the density (left) and current(right) during the $^{208}$Pb~+~$^{208}$Pb collision with $E_{\rm c.m.} =$~700~MeV and $b$~=~4~fm. The total angular momentum of the system is 236.9~$\hbar$.}
    \label{fig:Pb208_dens_cur}
\end{figure*}

\section{Method}

A  version of the LISE code \cite{Shi:2020} without superfluidity is used to solve the TDHF equations with the Runge-Kutta method at the order 4 with a time step $\Delta t$=0.3 fm/c. The evolution is described in a three-dimensional Cartesian grid discretized with $N_x=N_y=$~60, $N_z=28$, and a mesh spacing $dx=0.9$ fm. The  Sly4d functional~\cite{Kim:1997} is used to avoid the problem of the center of mass correction. The initial Hartree-Fock (HF) wave functions $\Phi_i({\bf r})$ are obtained with the Sky3d code \cite{Schuetrumpf2018}. 

Fragments are placed on the diagonal of the x-y plan to maximize the initial distance (see Fig.~\ref{fig:Pb208_dens_cur}). The boost intensity and direction are determined to obtain a given impact parameter~$b$ and center of mass energy~$E_{\rm c.m.}$. The impact parameter is chosen in the x-y plan to induce a positive orbital angular momentum $\Lambda_z=b\sqrt{2\mu E_{\rm c.m.}}$ in the z-orientation. With $\mu$ the reduced mass of the system.

To investigate the observables related to each fragment, the system is separated into two using the following method. First, the principal axis of deformation is determined. In the intrinsic framework, the two-dimensional projected density is fitted with a sum of two Gaussians. The position of the cutting plan dividing the system in two is determined as i) perpendicular to the principal axis and 2) pass by the minimum density of the adjusted function projected on the principal axis. This dividing plan determines the projector operator $ {\bf \Theta_F}$ on the half-space containing the fragment $\bf F$.

The total spin of the fragments is computed as    
\begin{widetext}
\begin{align}
   {\bf J}_{\bf F} & = \hbar \int d{\bf r}  \sum_i   
   \langle \Phi_i({\bf r}) |  \left( (\hat{\bf r} - {\bf r}_{\rm c.m.}^{\bf F}  ) \times ( \hat{\bf p} - {\bf p}_{\rm c.m.}^{\bf F} ) +\hat {\bf s} \right )  {\bf \Theta_F} | \Phi_i({\bf r})  \rangle , \label{eq:j_frag}
  \end{align}
  \end{widetext}
  with ${\bf r}_{\rm c.m.}^{\bf F}$ and ${\bf p}_{\rm c.m.}^{\bf F}$ the position and momentum of the center of mass of the fragments.  
  In the following, only the z-component of the spin/angular momenta is considered, and the z subscript is omitted. It should be noted that, in the present mean-field approach, only the one-body observable is taken into account, which means that the x and y components of the spins will always be zero.
   The present article will not investigate the spin probabilities with the projection method \cite{Bertsch:2019,Scamps:2023a} since i) it is more complex to implement and ii) it involves two- (and more) body correlations which have to be handled carefully in one-body-based models.  
  
The contact time is computed as the time between the two maxima of the acceleration of the fragments.  See Appendix \ref{app:contact_time} for more details about this method.

\section{Results}

To understand the mechanisms involved in angular momentum transfer, progressively complex systems are examined. First two symmetric systems between spherical doubly magic nuclei, $^{208}$Pb~+~$^{208}$Pb and $^{40}$Ca~+~$^{40}$Ca. Then $^{40}$Ca~+~$^{208}$Pb which introduce asymmetry. Then, a system that includes initial deformation $^{50}$Ca~+~$^{176}$Yb.

\subsection{ Symmetric heavy system $^{208}$Pb~+~$^{208}$Pb}

\begin{figure}[h]
    \centering
    \includegraphics[width=\linewidth]{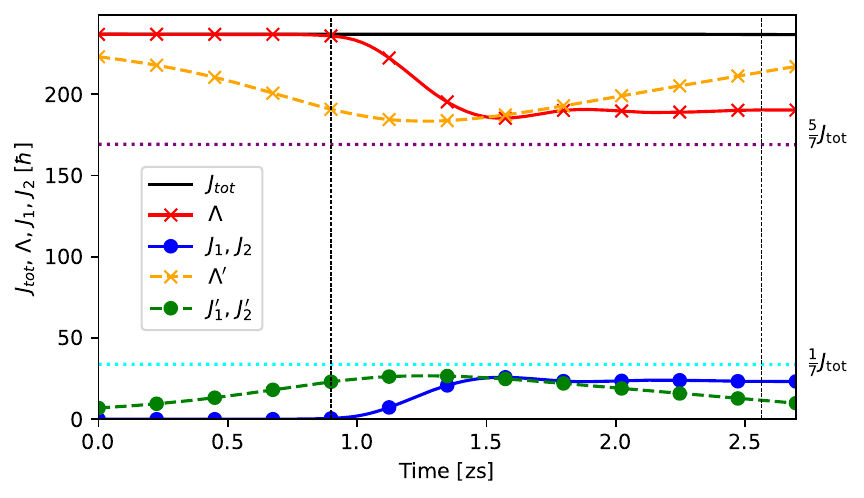}
    \caption{Evolution of the z-component of the total spin, relative orbital angular momentum, and spins of the fragments as a function of time for the reactions $^{208}$Pb~+~$^{208}$Pb at $E_{\rm c.m.}~=$~700 MeV and $b=4$~fm. The vertical dot lines show the moment of contact and scission as defined in Appendix \ref{app:contact_time}. The dashed and dotted lines represent different estimations of the equilibrium value with simple values of the moment of inertia and scission configuration (dotted lines) and from the TDHF moment of inertia (dashed lines) (see also text).}
    \label{fig:Pb208_AM_fct_t}
\end{figure}

To understand the mechanisms leading to spin transfer in reaction,  the first test case of the study is a symmetric reaction between spherical magic nuclei. The absence of transfer and fusion simplifies the analysis of the results. Fig.~\ref{fig:Pb208_AM_fct_t} shows the z-component of the spin of the fragments during a collision at a center of mass energy $E_{\rm c.m.}=$~700~MeV (25$\%$ above the required energy for the fragments to come into contact and for nuclear interactions to become significant\footnote{This energy is computed as the extrapolation of the fusion barrier formula of Ref. \cite{Scamps2018empirical}}) and an impact parameter $b$~=~4~fm.  As expected in the TDHF calculation, the total angular momentum,
\begin{align}
J_{\rm tot} = \Lambda + J_1 + J_2 
\end{align}
is well conserved.

On the $\Lambda_0~=~$236.9~$\hbar$ of the initial relative orbital angular momentum, only a fraction (44~$\hbar$) is transferred to the spin of the fragments. 
The main transfer takes place after the two fragments get in contact.
The relative orbital angular momentum is quickly transferred to the fragments in a time-scale of around 120~fm/c~=~0.4~zs. This value differs significantly from the 1~zs time-scale found in Ref. \cite{simenel2020timescales}. However, in Ref. \cite{simenel2020timescales}, the time scale is estimated from the contact time and not from the time evolution of the angular momentum. 
While the two fragments are still in contact the transfer stops at around $t$~=~1.5~zs and the spins are almost constant during the separation phase (see also Fig.~\ref{fig:Pb208_dens_cur}). In the following sections, the value of the spins at equilibrium will be further discussed.

\begin{figure}[h!]
    \centering
    \includegraphics[width=\linewidth]{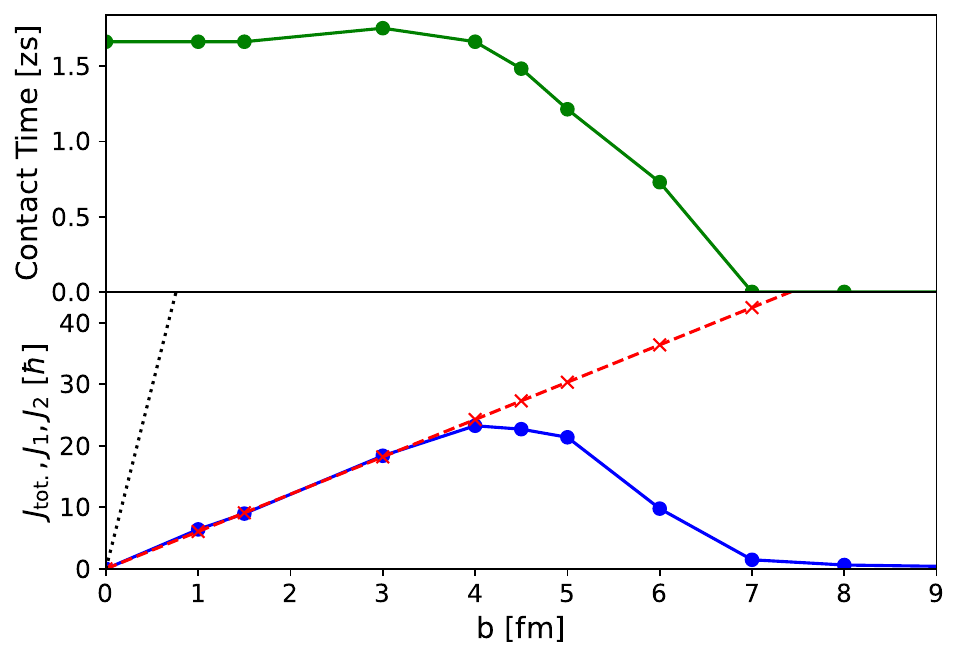}
    \caption{ The contact time (top) and final spin of the fragments (blue dots, bottom) are shown as functions of the impact parameter for the $^{208}$Pb~+$^{208}$Pb reaction at 700~MeV. In the bottom panel, the red dashed line represents $J=\alpha J_{\rm tot}$ (see text), with $J_{\rm tot}$ indicated by the black dotted line.}
    \label{fig:Pb208_fct_b}
\end{figure}

By varying the impact parameter at a fixed center of mass energy, reactions with different contact times can be investigated.
 Fig.~\ref{fig:Pb208_fct_b} shows the contact time and the transferred spin to the fragments as a function of the impact parameter. When the impact parameter is smaller than about 4-5~fm the fragments stay in contact for around 1.5~zs, while for larger impact parameters the fragments barely touch and separate quickly. 
As shown in the figure, the final angular momentum of the fragments is proportional to $\Lambda_0$ when the contact time is sufficiently long to allow equilibration. This equilibration occurs when the contact time is approximately 1.5 zs.

The transition between the equilibration to a linearly increasing equilibrium value and the aborted transfer with a lower and lower contact time create a curve with a maximum spin of 22~$\hbar$ for each fragment.  

The equilibration value can be understood with the sticking and rolling model \cite{Bass1980,Tsang1974}. In this model, it is assumed that the fragments are rigid and spherical. Friction plays a role in the contact zone of the two fragments. A first equilibrium is reached when the difference of current between the fragments due to the rotation is zero at the contact point. This equilibrium is achieved with the sliding friction force. This leads to the rolling limit when $\omega r = \omega_1 R_1 +  \omega_2 R_2 $. With $\omega$, $\omega_i$ the angular velocities of the system and the fragments, $r$ is the distance between the fragments, and $R_i$ is the radius of the fragment $i$. In an asymmetric reaction, the light fragment can roll around the heavy fragment, which induces another friction around the neck. This rolling friction is expected to be weaker by one order of magnitude than the sliding friction~\cite{Tsang1974}.
The final equilibrium is achieved when all angular velocities are equal and correspond to a di-nuclear system quasi-static in its rotational frame.
This condition is satisfied when:
\begin{align}
\frac{J_1}{I_1} =  \frac{J_2}{I_2} =  \frac{\Lambda}{I_{\Lambda}} \label{eq:equal_ang_vel}
\end{align}
which implies a proportional relationship between the spin of each fragment and the initial total angular momentum:
\begin{align}
J_i'  = \frac{I_i}{I_{\Lambda}+I_1+I_2}  J_{\rm tot} \label{eq:model}
\end{align}
Here, $I_i$ represents the moment of inertia of the fragments and $I_{\Lambda} = \mu r^2$ the relative moment of inertia. In a simplified version, assuming a moment of inertia $I_i=\frac25 M_i R_i^2$, and a scission radius $r=R_1+R_2$, for a symmetric system the equilibration angular momentum should be $J_i=\frac17$ and $\Lambda=\frac57$. These values shown in Fig.~\ref{fig:Pb208_AM_fct_t} by dotted lines are not far from the TDHF results. A simple improvement of this model can be obtained by assuming a scission distance $r=R_1+R_2+3.5$ fm. The value of 3.5~fm is adjusted to reproduce the TDHF results. This leads to the coefficient of proportionality
\begin{align}
\alpha = \frac{I_i}{I_{\Lambda}+I_1+I_2} \simeq 0.102, 
\end{align}
which determines the red dashed curve in Fig.~\ref{fig:Pb208_fct_b}. A more coherent way to test the assumption of equal angular velocity of eq. \eqref{eq:equal_ang_vel} is to determine the rigid moment of inertia from the TDHF density. The resulting expected equilibration values for the relative orbital angular momentum $\Lambda'$ and spin of the fragments $J_i'$ are shown in Fig.~\ref{fig:Pb208_AM_fct_t} with dashed lines. It can be seen that the angular momentum of the fragment reaches the equilibrium value and stick to it when it starts to decrease. Then, the angular momentum keeps almost constant after t=2~zs while the scission is not yet reached. 

 The density and currents for that reaction are shown in Fig.~\ref{fig:Pb208_dens_cur}. The friction leading to a rotation of the fragment is visible in particular between panels at time t=0.99 to 1.65 zs. Just before friction plays a role at time 0.99 zs, the currents are non-zero around the neck showing a confrontation between the velocities of each sphere. The current at the neck disappears progressively until complete equilibration around t =1.6 zs.
 
Although the simple model of sticking condition does not take into account the neck formation and the quantal nature of the rotation which are ingredients present in the TDHF approach, the model still quantitatively explains the friction equilibration found in the microscopic description.

\subsection{Short contact time, $^{40}$Ca~+~$^{40}$Ca}

\begin{figure}[h!]
    \centering
    \includegraphics[width=\linewidth]{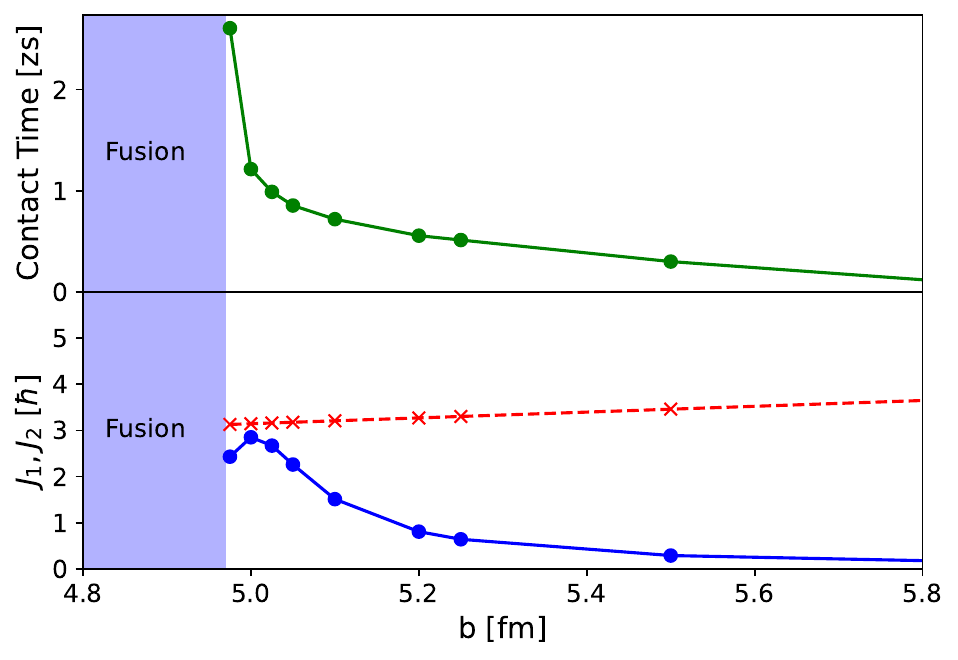}
    \caption{ Same as Fig.~\ref{fig:Pb208_fct_b}  for the $^{40}$Ca~+~$^{40}$Ca at  $E_{\rm c.m.}$~=~70~MeV. }
    \label{fig:Ca40_fct_b}
\end{figure}

In the case of $^{40}$Ca~+~$^{40}$Ca, with a center of mass energy of 70~MeV (1.55 times the fusion barrier), small impact parameters result to the fusion of the system due to the lighter fragments.  
Then, the angular momentum transfer reaction takes place only when there is a re-separation, i.e. when the effective barrier is higher than the center of mass energy. In the present case for $b$~>~4.97 fm. Except for $b$ values near the threshold, the contact time is short, lasting less than 1~zs as shown in Fig.~\ref{fig:Ca40_fct_b}. 

\begin{figure}[h!]
    \centering
    \includegraphics[width=\linewidth]{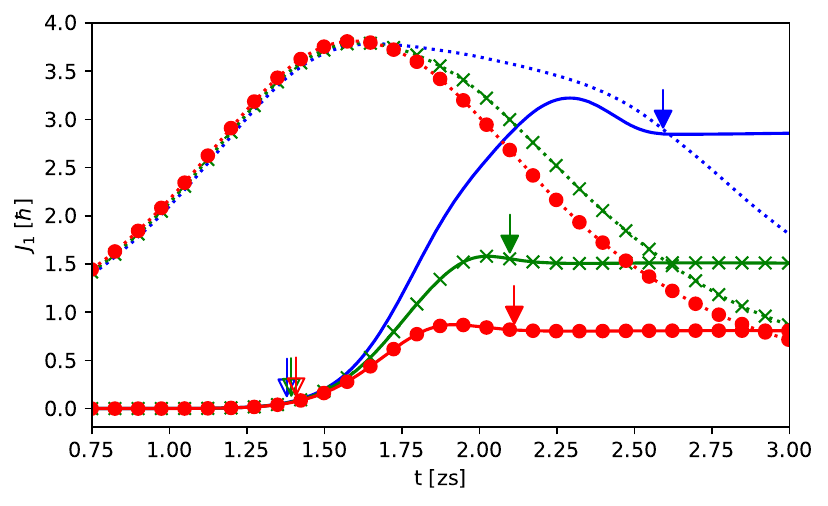}
    \caption{  The evolution of the fragment spin (solid line) is shown for the $^{40}$Ca~+~$^{40}$Ca reaction at $E_{\rm c.m.} =$~70~MeV with initial impact parameters of $b$~=5, 5.1, and 5.2~fm, represented by no symbol, crosses, and dots, respectively. The dotted line indicates the expected value from Eq. \eqref{eq:model}, calculated using the TDHF rigid moment of inertia. The open and filled arrows mark the contact and scission times, respectively.}
    \label{fig:AM_fct_time_Ca40}
\end{figure}

While the total spin is around 40 $\hbar$,  the maximum spin observed in the fragments is around 3 $\hbar$ for $b$ around the threshold. This value is significantly smaller than in the $^{208}$Pb~+~$^{208}$Pb case. The coefficient $\alpha$, computed assuming a scission distance $r=R_1+R_2+4$ fm, is now around 0.071, which is much smaller to the simplified model value of $\frac17$ due to the relative importance of the neck compared to the radii of the fragments.

Fig.~\ref{fig:AM_fct_time_Ca40} shows the time evolution of the spin of the fragments for different impact parameters. It shows that the rate of angular momentum transfer depends on the impact parameter. For $b$=5 fm, the maximum transfer rate of angular momentum is 6~$\hbar$/zs and about 2.3~$\hbar$/zs for b=5.2 fm. This figure also shows that equilibrium is achieved for b=5 fm after almost 1 zs of contact. Just before the scission, there is a small decrease following the evolution of the expected equilibrium value.
In the case of $b$~>~5 fm the spin transfer is too slow and the contact time too short to achieve a full equilibrium of the system's rotation.

The different transfer rates can be understood simply by assuming that the tangential friction depends on the overlap between the two densities as in Ref. \cite{Tsang1974}. Then a larger impact parameter will lead to less tangential friction.

\begin{figure}[h!]
    \centering
    \includegraphics[width=\linewidth]{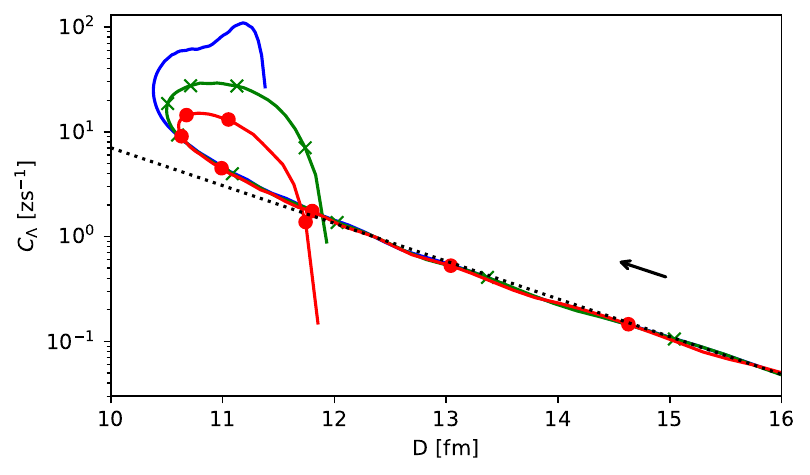}
    \caption{ Friction coefficient as a function of the distance between the two $^{40}$Ca fragments. The same symbols are used in Fig.~\ref{fig:AM_fct_time_Ca40}. The dotted line represents a fit at a large distance with an exponential (see text). The arrow shows the direction of the motion during the approach phase.}
    \label{fig:friction_coef_Ca40}
\end{figure}

Assuming a similar parametrization of the macroscopic equation of motion of the relative orbital angular momentum than~\cite{Wash2020},
\begin{align}
\frac{d\Lambda}{dt} = - C_\Lambda \frac{\left( \Lambda - \Lambda_{eq} \right)}{\mu},
\end{align}
where one replace $\Lambda$ by $\Lambda - \Lambda_{eq}$ since it was assumed in Ref.~\cite{Wash2020} that the equilibrium value of the relative orbital angular momentum was zero. With  the equilibrated value for the relative orbital angular momentum,
\begin{align}
\Lambda_{eq} = \frac{I_{\Lambda}}{I_{\Lambda}+I_1+I_2}  J_{\rm tot}. \label{eq:model_Lambda}
\end{align}
The deduced  tangential friction coefficient,
\begin{align}
C_\Lambda = -  \frac{\mu}{\left( \Lambda - \Lambda_{eq} \right)} \frac{d\Lambda}{dt},
\end{align}
 is shown in Fig.~\ref{fig:friction_coef_Ca40}. For distances larger than 12~fm, the coefficient varies exponentially with the distance. Then, it can be well-fitted by a function,
 \begin{align}
 C_\Lambda (D) \simeq e^{C_0 D + C_1},
 \end{align}
with $C_0$~=~-0.83~fm$^{-1}$ and $C_1$~=~10.25.

At smaller distances, the neck formation significantly modifies the picture with an increase in the coefficient as a function of time.
This behavior shows the limitation of the macroscopic models that neglect the dynamic aspect of the neck formation. 
Note that the curves are not shown fully during the separation phase, because during that phase other effects can slightly change the value of the relative orbital angular momentum leading to values of $C_\Lambda$ that can not be interpreted as a friction coefficient.

\subsection{Asymmetric reaction, $^{40}$Ca~+~$^{208}$Pb}

Asymmetry in reactions induces two complications in the mechanism of angular momentum transfer. First, as seen above, there are two types of friction, rolling and sticking with their associated time scale. Second, the asymmetry allows for a transfer of nucleons \footnote{Transfer can also happen in symmetric reactions but will not affect the one-body variables in mean-field calculations}. The transfer is expected to induce an internal angular momentum in each fragment \cite{Dossing:1985} and also change the moment of inertia during the contact phase.

The spin of each fragment after the collision, as determined by the TDHF simulation for the reaction $^{40}$Ca~+~$^{208}$Pb at $E_{\rm c.m.}$~=~200 MeV (16\% above the barrier), is shown in Fig.~\ref{fig:Ca40_Pb208}. The evolution of the two spins as a function of the impact parameter is more complex than in the previous case for the following reasons:

\begin{itemize}
\item As shown in the first panel, the direction of proton transfer changes depending on the impact parameter.
\item The second panel indicates that the contact time now reaches values much longer than in the two previous cases.
\item The transferred spin must now be distributed between the fragments. As seen in the third and fourth panels, the spins of both fragments do not exhibit a simple evolution.
\end{itemize}

The different aspects of the reactions will now be investigated in the following sections.

\subsubsection{Sliding and rolling friction}

\begin{figure}[h!]
    \centering
    \includegraphics[width=\linewidth]{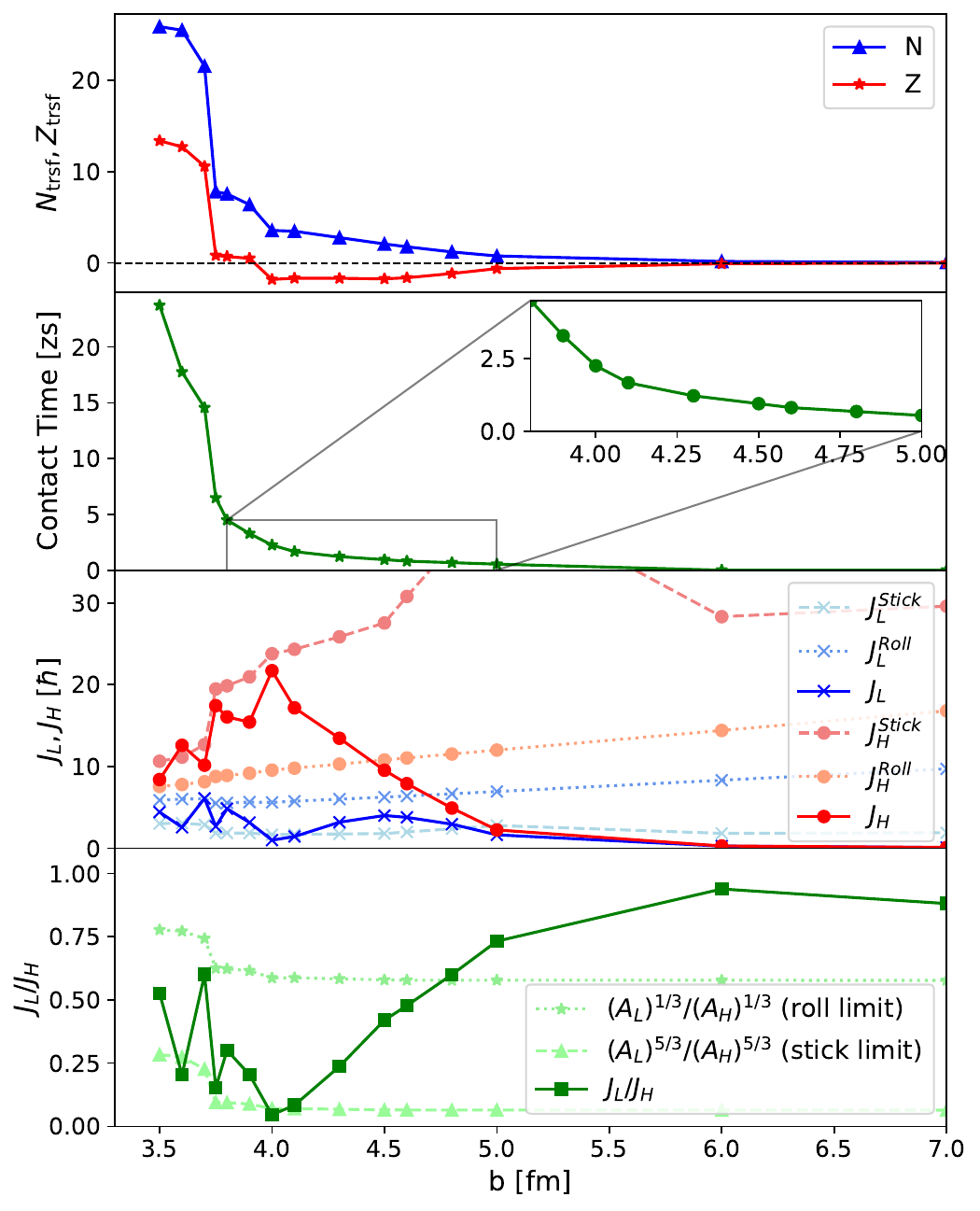}
    \caption{Results of the $^{40}$Ca~+~$^{208}$Pb collision at an energy of 200 MeV. The first panel shows the number of particles transferred from the heavy fragment to the light one. The second panel shows the contact time between the two fragments. The light (heavy) spin $J_1$ ($J_2$) is shown on the third panel and compared to the sticking and rolling limit assuming no transfer. The last panel compares the ratio between the two spins with the expected value assuming rolling or sticking dissipation (see text).}
    \label{fig:Ca40_Pb208}
\end{figure}

To determine the rolling and sticking equilibrium, one can take advantage of the fact that they predict different values of the ratio $\frac{J_1}{J_2}$ \cite{Bass1980} at equilibrium,
\begin{align}
 \frac{J_1}{J_2} = \frac{R_1}{R_2} \simeq \frac{A_1^{1/3}}{A_2^{1/3}} \quad &\text{Rolling equilibration} \\
 \frac{J_1}{J_2} = \frac{I_1}{I_2} \simeq \frac{A_1^{5/3}}{A_2^{5/3}} \quad &\text{Sticking equilibration} 
\end{align}
In the case of $^{40}$Ca~+~$^{208}$Pb, without transfer that ratio $J_L/J_H$ is 0.58 (0.06) for the rolling (sticking) equilibration.

Let's start the analysis with the large impact parameter $b$>5 fm. For these impact parameters, the two fragments barely touch each other, the transfer is negligible and the contact time is short. Only a small overlap between the two densities is at play to induce tangential friction. In this case, the ratio  $\frac{J_1}{J_2} $ is  closer to one than the ratio $\frac{A_1^{5/3}}{A_2^{5/3}} $. This can be understood by the fact that in the model of Ref. \cite{Tsang1974}, the parameters $g_1$ and ($g_2$) which represent the distance between the overlap region and the center of the fragments 1 (2), are assumed to be close to $R_1$ ($R_2$). If one assumes a fixed distance $d$ between the surface of the fragments and the overlap region, the ratio  $\frac{J_1}{J_2} \simeq \frac{g_1}{g_2} = \frac{R_1+d}{R_2+d} $ is closer to 1 than $\frac{A_1^{1/3}}{A_2^{1/3}}$ which is consistent with the microscopic simulation.

For impact parameters between 5 fm and 4 fm, a competition between the rolling and sticking friction takes place. With longer and longer contact time, the ratio $\frac{J_1}{J_2}$ decreases and reaches the expected sticking equilibrium around $b$~=~4~fm with a contact time of about 2~zs. 
Before that equilibrium, the light fragment spin passes by a maximum.  Indeed for the light fragment, the rolling equilibrium value is higher than in the sticking condition. That maximum is obtained for $b$~$\simeq$~4.5 fm, which corresponds to a contact time of about 1~zs. 
Between 4 and 5~fm, a transfer of a few neutrons and protons takes place but appears to not affect the spin of the fragments. 
Although this transfer may complicate the present picture, one can estimate the rolling relaxation time of about 1~zs and the sticking one of about 2~zs. Which induces roughly a factor of two between the two friction coefficients while macroscopic calculations estimated the rolling friction to be 10 times smaller than the sliding one \cite{Bass1980}.

\begin{figure}[t]
    \centering
    \includegraphics[width=\linewidth]{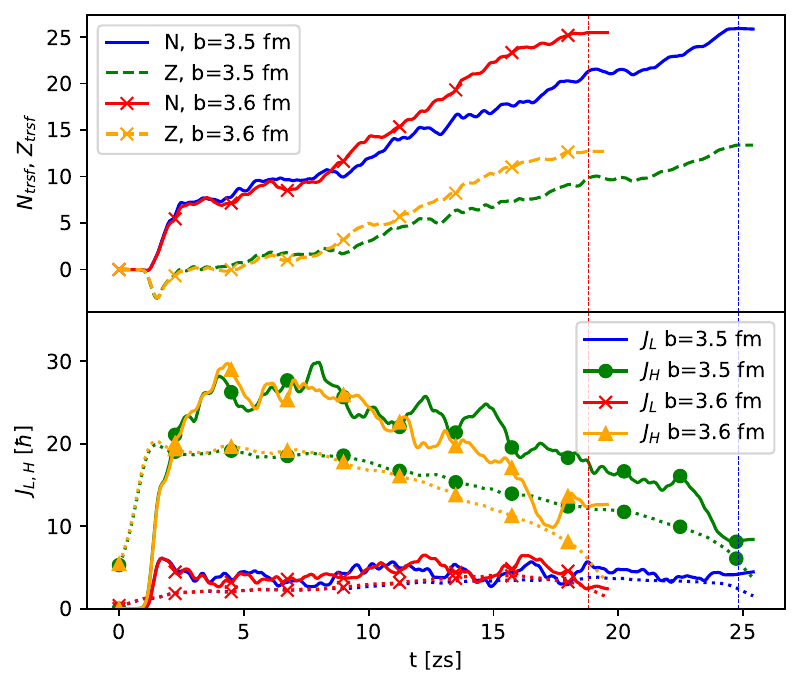}
    \caption{ Top: Transfer of nucleons from the heavy to the light fragments. Botton: Evolution of the spin of the fragment (solid line) for the  $^{40}$Ca~+~$^{208}$Pb reaction at $E_{\rm c.m.} =$ 200 MeV for different initial impact parameters. The dashed lines correspond to the expected value of eq. \eqref{eq:model} with the rigid moment of inertia. The vertical dashed line shows the scission time. }
    \label{fig:AM_fct_time_sev_b_Ca40_Pb208}
\end{figure}

The long contact time trajectories with impact parameters between 3.5 fm and 4 fm are more complex with a stochastic behavior and a non-negligible effect of the transfer. First, the transfer changes the equilibrium value. The mass equilibrium is expected to take place with a timescale of the order of 20 zs  \cite{simenel2020timescales}.  Note that the mass equilibrium does not mean a symmetric splitting, since shell effects can induce an asymmetric splitting \cite{Simenel2021} similar to the mass asymmetric fission \cite{Scamps:2018}. As shown on the bottom panel of Fig.~\ref{fig:Ca40_Pb208}, the transfer equilibrium brings both ratios closer to 1.  However, the TDHF results alternate between different values.  To understand this behavior, Fig.~\ref{fig:AM_fct_time_sev_b_Ca40_Pb208} shows the transfer of nucleons and spin as a function of time for impact parameter $b$~=~3.5 and 3.6~fm. The evolution of the two systems is similar for the first 10 zs and then starts to deviate. The final spin of the fragments is quite different, while the expected value of the sticking condition is similar. This is due to the stochastic evolution of the spin of the fragments. Because of the large excitation energy created by the equilibrium of the masses, a large variety of modes is excited in the fragments leading to large oscillations of their spins.
Then, the random nature of the breaking of the neck induces fluctuations in the final value of the fragment's spin.

\subsubsection{Direct effect of particle transfer}

\begin{figure}[h!]
    \centering
    \includegraphics[width=\linewidth]{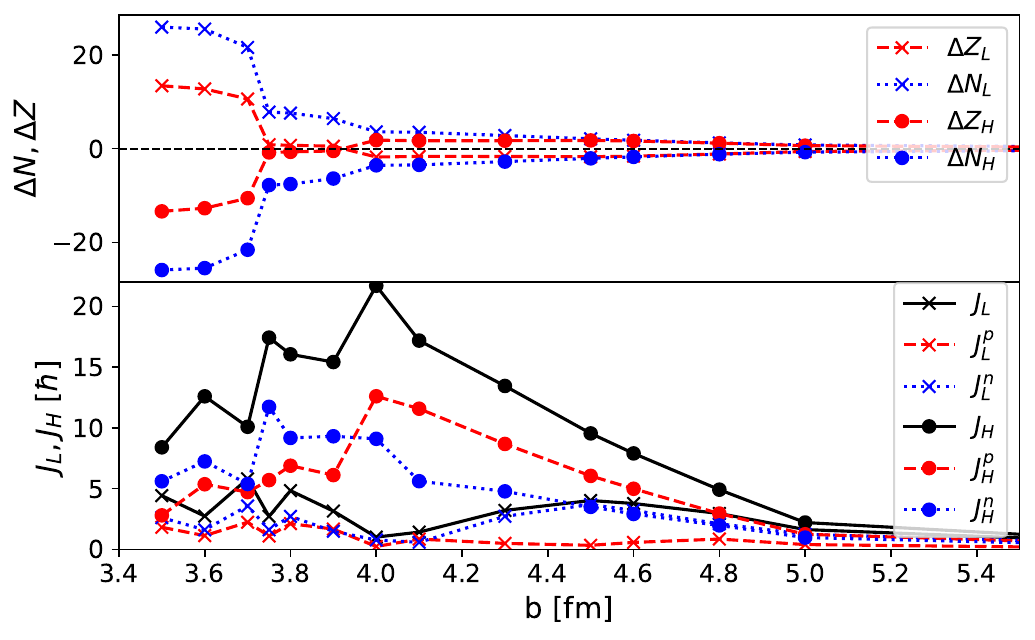}
    \caption{Top: Variation of the number of neutrons (dotted lines) and protons (dashed line) in light (dots markers) and heavy (cross markers) fragments for the  $^{40}$Ca~+~$^{208}$Pb reaction at $E_{\rm c.m.} =$ 200 MeV. Bottom: total spin in light (dots markers) and heavy (cross markers) fragments, the spin computed only from protons (dashed line) and neutrons (dotted lines) is also shown. }
    \label{fig:Ca40_Pb208_np}
\end{figure}

The isospin mechanism can play a significant role in reactions \cite{godbey2017dependence}. As discussed in Ref.~\cite{Sekizawa2013}, the direction of proton transfer changes as a function of the contact time due to neck formation.

Figure \ref{fig:Ca40_Pb208_np} illustrates another aspect of the results described in the previous section: the isospin dependence of the spin. In an approach assuming rigid fragments, a fixed ratio between the spins of neutrons and protons would be expected.
\begin{align}
\frac{J_i^n}{J_i^p} = \frac{N_i}{Z_i}
\end{align}
This is not what is found in Fig.~\ref{fig:Ca40_Pb208_np}. For example, for $b$~=~4.5~fm, the spin of the light fragment is almost only carried out by the neutrons, and in the heavy fragment, the spin of protons is higher than for neutrons. For $b$~<~4~fm the proton transfer changes direction and the ratio between the neutron and proton spin gets closer to  $N/Z$. The transition takes place at the same impact parameter ($b$~<~4~fm) as the change of direction of the proton transfer.

This result suggests that i) the fragment's rotation is highly non-rigid and ii)  when nucleons are transferred they bring angular momentum. At $b$~=~4.5~fm, when protons are transferred from the light to the heavy fragments, there is an excess of spin for protons in the heavy fragment and a depletion in the light fragment. For $b$~<~4~fm, this isospin effect is attenuated since the transfer of neutrons and protons is in the same direction.

\subsubsection{Spin vs mass equilibration}

\begin{figure}[t]
    \centering
    \includegraphics[width=\linewidth]{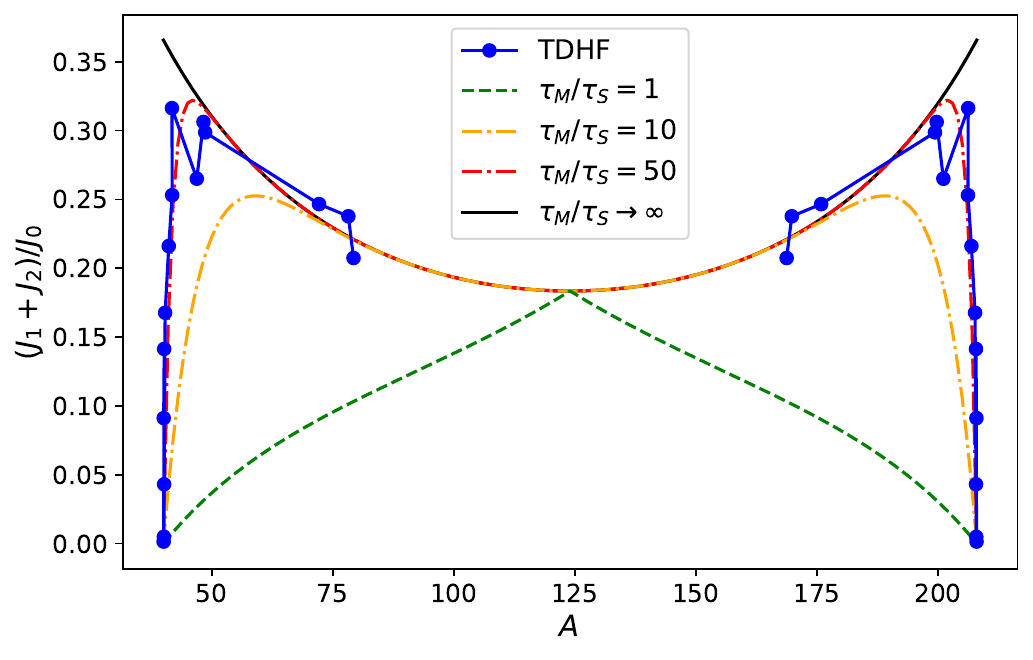}
    \caption{ Total intrinsic spin of the fragments as a function of the mass for the reaction $^{40}$Ca~+~$^{208}$Pb at $E_{\rm c.m.}$~=~200~MeV. The TDHF results are compared to the simpler model with eq. \eqref{eq:relax}.}
    \label{fig:Ca40_Pb208_Jsum_fct_A1_A2}
\end{figure}

 \begin{figure*}[t]
    \centering
    \includegraphics[width=.99\linewidth]{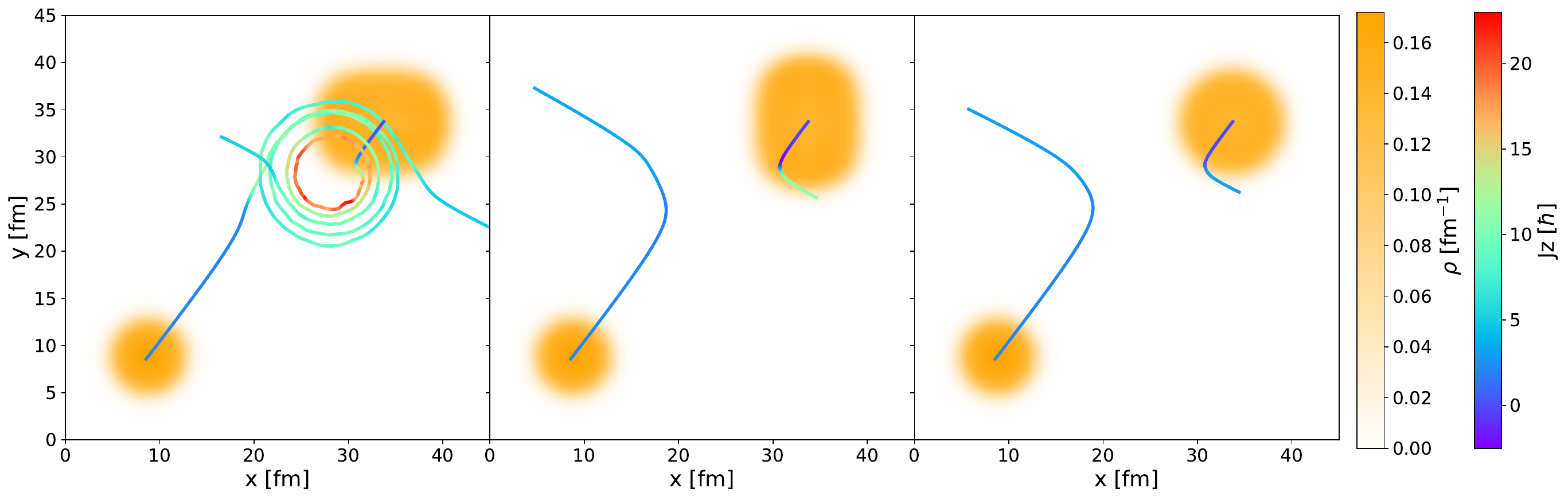}
    \caption{Trajectories of the $^{50}$Ca~+~$^{176}$Yb reaction in the x-y plane for an impact parameter $b$~=~3.8~fm, are shown in three subfigures, with initial orientations of x (left), y (middle), and z (right). The background displays the initial density, and the color of each trajectory line represents the magnitude of the fragment's spin.}
    \label{fig:density_with_trajectory_at_t0}
\end{figure*}

Introduced in Ref.~\cite{Back1990}, the spin vs mass figure (Fig.~\ref{fig:Ca40_Pb208_Jsum_fct_A1_A2}) shows the relative speed of the spin and mass equilibration. Assuming simple relaxation time $\tau_M$ and $\tau_J$ respectively of the mass and spin, the following relation is found~\cite{Back1990},  
\begin{align}
J&=J_1+J_2 \nonumber \\
&=J_{eq}(A_1,A_2) \left[  1 - \left(    \frac{|A_{eq}-A|}{A_{eq}}-A_0  \right)^{\tau_M/ \tau_J}   \right] \label{eq:relax}
\end{align}
With the sticking equilibration $J_{eq}(A_1,A_2)$ as the sum of the fragments $J_i'$ of eq.~\eqref{eq:model} assuming a scission at a distance $r=R_1+R_2+4$~fm, the estimated spin-mass curve is shown on the figure with different ratio between the relaxation time.  

The value $\tau_M/ \tau_J$~=~50 appears to fit correctly the TDHF results. This is higher than the ratio of 20 found in Ref.~\cite{simenel2020timescales}, which also use the TDHF model, however, here only one reaction is taken into account and a partial equilibration is found after 20~zs, which is similar to the time needed for equilibration in \cite{simenel2020timescales}. Nevertheless, this ratio is much higher than the one adjusted on experimental data in Ref.~\cite{Back1990} with a value between 2 and 3. Also, the experimental curve has an n shape while the results of Fig.~\ref{fig:Ca40_Pb208_Jsum_fct_A1_A2} present a U shape. This discrepancy can be attributed to the fact that the spin is not directly measured in experiments; instead, the $\gamma$ multiplicity is measured.

\subsection{Effect of the deformation, $^{50}$Ca~+~$^{176}$Yb}

In the case of the $^{50}$Ca~+~$^{176}$Yb reaction, the $^{176}$Yb is initially deformed with a deformation parameter of $\beta_2$~=~0.196, with the definition of \cite{Schuetrumpf2018}.  The calculation is done here with a center of mass energy of 172~MeV (16~$\%$~above the barrier) for various impact parameters and initial orientations. This is the same system as in Ref. \cite{ Simenel2021} where it was shown that the large transfer is affected by the same shell effect as in fission.

\begin{figure}[h!]
    \centering
    \includegraphics[width=\linewidth]{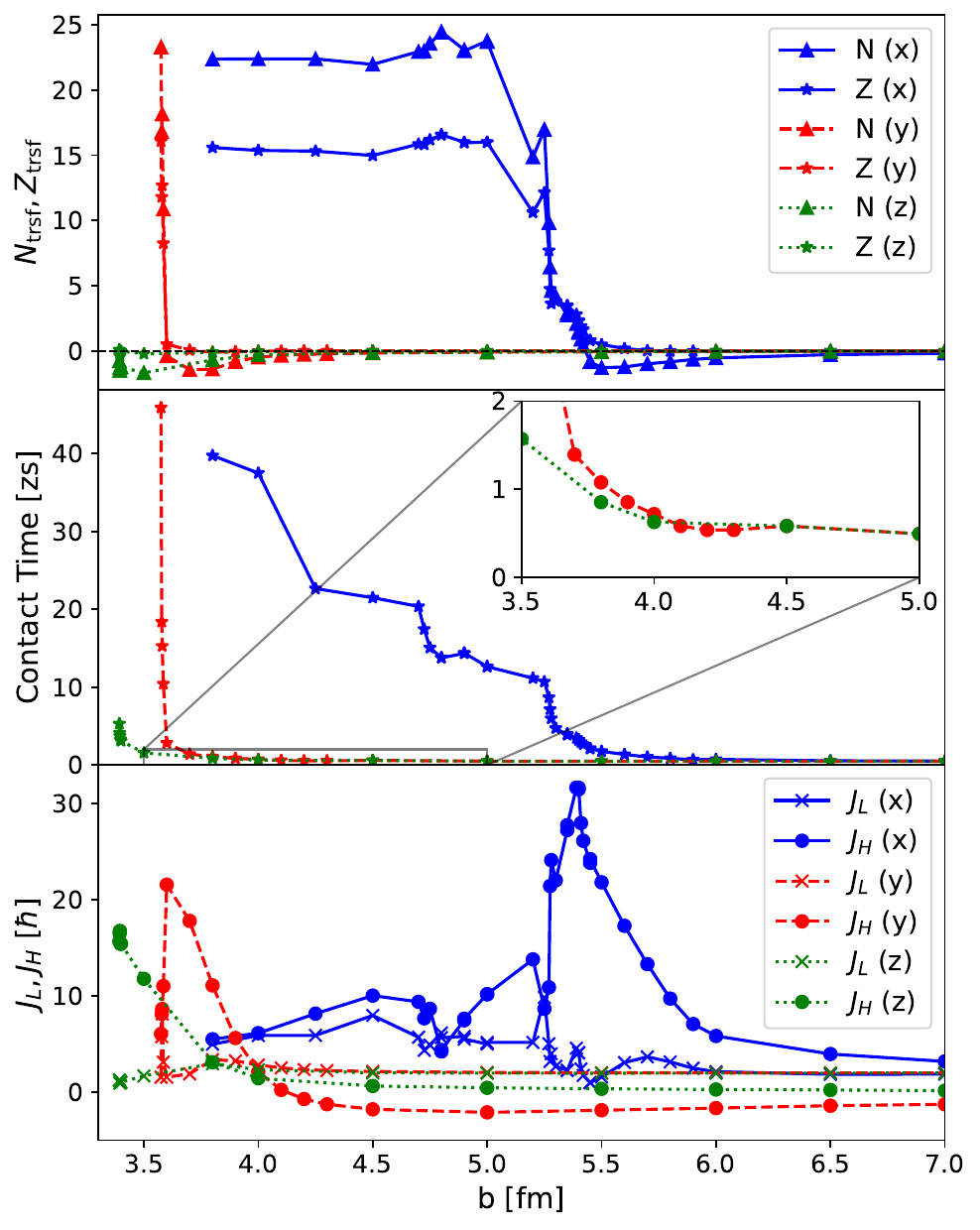}
    \caption{Transfer of nucleons, contact time, and spin of the final fragments as a function of the impact parameter respectively on the top, medium, and bottom panels for the $^{50}$Ca~+~$^{176}$Yb reaction. In each cases, fusion occurs for impact parameters smaller than the lowest value of $b$ displayed. }
    \label{fig:Ca50_Yb176}
\end{figure}

Deformation introduces new complexities in spin transfer, including (i) the Coulomb torque~\cite{simenel2004quantum}, which can cause fragments to rotate before contact, (ii) the nuclei-nuclei force that can also induce a torque, and (iii) a strong dependence of the friction coefficient on the initial orientation because it modifies the geometry of the system during the contact of the two fragments. While the results are inherently dependent on the entire 2D space of $\theta$ and $\varphi$ angles for the $^{176}$Yb fragments, for simplicity, only three initial configurations (x, y, and z) are considered here, as illustrated in Fig.~\ref{fig:density_with_trajectory_at_t0}.  It should also be noted that the initial $^{50}$Ca fragment has an initial spin in the z-direction of 2 $\hbar$ due to the open-shell nature of this one in the framework of the Hartree-Fock theory.  This small spin initial value in an arbitrary direction is spurious and will be neglected in the following discussion since it will not affect significantly the mechanisms described in this section.

Figure \ref{fig:density_with_trajectory_at_t0} shows the trajectory of the fragments for each orientation. As expected the initial configuration of the Yb fragment changes drastically the dynamics of the collision since for the x-orientations the two fragments get into contact and so the nucleon-nucleon interaction plays a role resulting in prolonged contact time and substantial transfer.  In the y- and z-orientations the Coulomb repulsion dominates and leads to a quick separation of the fragments. Note that for smaller values of $b$ the fragment come in contact for all initial orientations.

Notably, in the initial y-direction case, the spin component of the heavy fragment along the z-axis starts negative before becoming positive. This behavior is driven by the initial influence of the long-range Coulomb torque, which, in this configuration, induces a negative angular momentum. Subsequently, the nucleus-nucleus interaction generates a stronger torque in the opposite direction, rapidly leading to a positive $Jz$ value.
This negative value of the spin, which induces an increase of the relative orbital angular momentum, i.e. a spin transfer in the opposite direction as the natural one, is visible also in Fig.~\ref{fig:Ca50_Yb176}, for impact parameters above 4.5~fm, the final spin of the heavy fragment is about -2~$\hbar$. For the x-direction, the Coulomb torque induces a positive value of the spin of the heavy fragment. This effect is visible for $b$~>~6~fm since for smaller values the spin is strongly affected by the nucleus-nucleus interaction and friction.

Beyond the effect of the Coulomb torque, the impact of the orientation is highly significant as shown in Fig.~\ref{fig:Ca50_Yb176}. Deep inelastic types of collision characterized by a large transfer and long contact time arise for $b\simeq$~5.5 fm, for the x-orientation, while much smaller impact parameters are required for y- and z-orientation, respectively $b\simeq$~3.7 fm and $b\simeq$~3.4 fm. However, the threshold impact parameter at which the system does not re-separate is much less affected by the orientation with a value from 3.39~fm for the z-orientation to 3.7~fm for the x-orientation. This allows for a wide range of impact parameters for deep-inelastic collisions in the initial x-orientation. 

 For the x-orientation, the evolution of the spin is similar to the $^{40}$Ca~+~$^{208}$Pb results shown in Fig.~\ref{fig:Ca40_Pb208}, with a rolling equilibration around $b$~=~5.7~fm associated with a peak of the light fragment spin. The sticking condition equilibration appears to be fulfilled for $b\simeq$~5.4~fm, leading to a peaked value of 30~$\hbar$ for the heavy fragment. Then for larger contact time, transfer and excitation energy lead to values of the spin between 5 and 10 $\hbar$ with large stochastic fluctuations.

\begin{figure}[t]
    \centering
    \includegraphics[width=\linewidth]{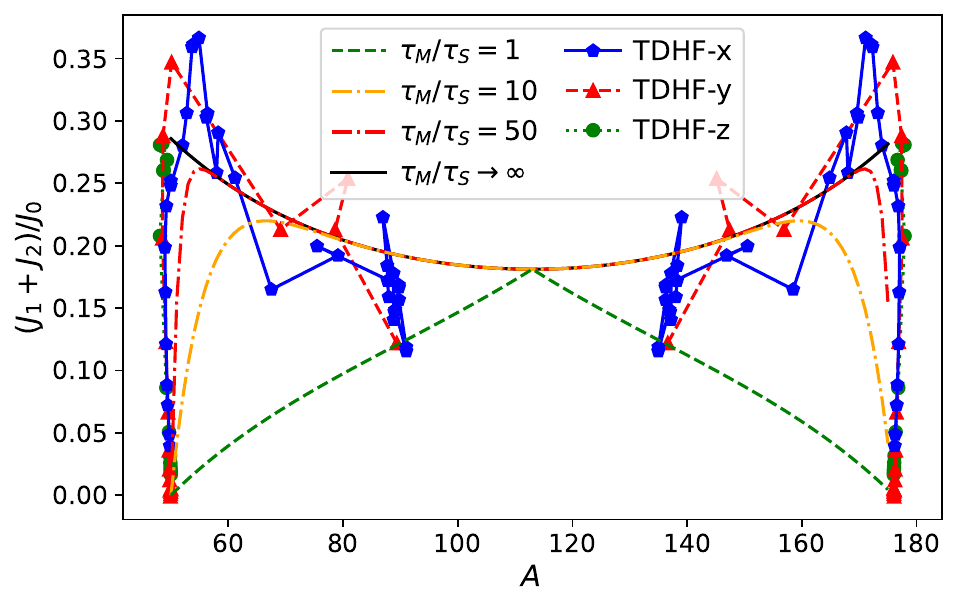}
    \caption{ Same as Fig.~\ref{fig:Ca40_Pb208_Jsum_fct_A1_A2}  for the $^{50}$Ca~+~$^{176}$Yb reaction with different initial orientations.}
    \label{fig:Ca50_Yb176_Jsum_fct_A1_A2}
\end{figure}
 
 Fig. \ref{fig:Ca50_Yb176_Jsum_fct_A1_A2} shows that there is no value of the ${\tau_M/ \tau_J}$ ratio that can reproduce the total spin versus mass curve. 
 It is then not possible to establish an estimation of the time scale of the mass and spin equilibration from such an experimental curve. The main reason here is that the mass equilibration starts after the equilibration of spin.



\section*{Conclusion}

To summarize, the transfer of spin between the initial relative orbital angular and the intrinsic spin of the fragments is studied extensively in several nuclear reactions with the TDHF theory. Here, the different conclusions about the mechanisms leading to a transfer of spin are listed,

\begin{itemize}

\item For deformed fragments, the Coulomb torque at large distances can induce a spin in the fragments that can either increase or decrease the relative orbital angular momentum depending on the initial orientation of the fragments. At short distances, the nucleus-nucleus interaction should also generate a torque. However, both effects have less influence on the final spin values than tangential dissipation

\item A direct effect of nucleon transfer bringing their angular momentum to the other fragment, that is visible on the isospin dependence of the spin of the fragments.

\item Transfer of nucleons can also play an important role in modifying the moment of inertia of the fragments and so the equilibrium values. This effect was neglected in previous macroscopic studies.

\item While it's well-established that both sliding and rolling friction contribute to tangential dissipation, the presence of the neck significantly alters the spin evolution of the fragment. Unlike earlier macroscopic calculations, which estimated sliding friction to be an order of magnitude larger, present microscopic calculations suggest that sliding friction is roughly twice the rolling friction coefficient. This behavior is attributed to the neck, which quickly locks the fragments into alignment with the reaction axis.

\item When the contact time is long, the excitation energy induces large stochastic fluctuations of the spins of the fragments.

\item For the reasons shown in the previous points, in contradiction to previous studies, the spin of the fragments does not always increase with time. Then the present calculations do not support the idea that the spin of the fragments could be used as a "clock", for example, to distinguish between the deep-inelastic collisions and fusion-fission.

\end{itemize}

Although a large change in the presented results is not expected with the inclusion of pairing, it would be interesting to investigate the effect of pairing on the sticking and rolling friction in the future. This could be explored using modern Time-dependent Hartree-Fock-Bogoliubov codes \cite{Shi:2020,hashimoto2016gauge,magierski2017novel}.

\begin{acknowledgements}
I gratefully acknowledge support from the CNRS/IN2P3 Computing Center (Lyon - France) for providing computing and data-processing resources needed for this work.  This work was also granted access to the HPC resources of IDRIS under the allocation 2024-AD010515531 made by GENCI.
\end{acknowledgements}

\appendix 

\section{Contact time}
\label{app:contact_time}

The definition of the contact time often relies on a minimum value of the density at the neck. Here, a new definition is proposed based on the dynamical definition of scission in Ref. \cite{Scamps:2023} which was based on the maximum radial acceleration of the fragments.
\begin{align}
a_r = \frac{d^2 r(t)}{dt^2}
\end{align}
with $r$ the distance between the two fragments.
 This corresponds to the time when the Coulomb repulsion dominates the nuclear attraction and radial friction. 

\begin{figure}[h]
    \centering
    \includegraphics[width=\linewidth]{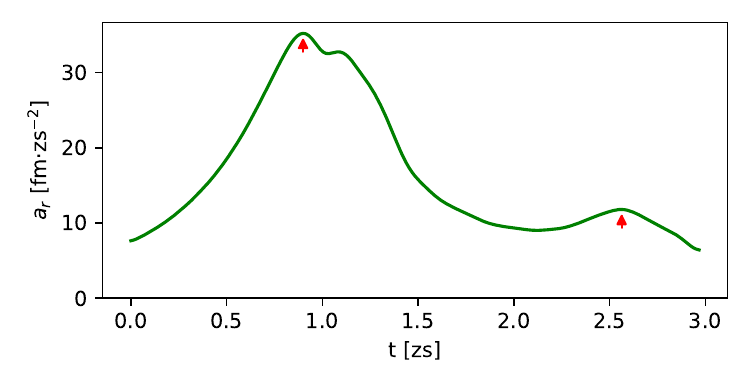}
    \caption{Radial acceleration of the fragments as a function of time, for the collision shown in Fig.~\ref{fig:Pb208_dens_cur}.}
    \label{fig:acceleration_vs_time_Pb208}
\end{figure}

An example of the application of the method is proposed in Fig.~\ref{fig:acceleration_vs_time_Pb208} with the radial acceleration as a function of time for the trajectory shown in Fig.~\ref{fig:Pb208_dens_cur}. The time of contact and separation are shown by arrows.


\providecommand{\selectlanguage}[1]{}
\renewcommand{\selectlanguage}[1]{}

\bibliography{local_fission.bib}

\end{document}